\documentclass[10pt,conference]{IEEEtran}
\IEEEoverridecommandlockouts
\usepackage{cite}
\usepackage{amsmath,amssymb,amsfonts}
\usepackage{algorithmic}
\usepackage{graphicx}
\usepackage{textcomp}
\usepackage{xcolor}
\usepackage{physics}
\usepackage{array}
\usepackage{comment}
\usepackage{url}
\usepackage{booktabs}
\usepackage{multirow}
\usepackage[countmax]{subfloat}
\usepackage{subfig}

\begin{document}

\title{Experimental Protocol Fingerprinting in Quantum Networks via Physical Layer Side Channel Analysis\\\

\vspace{-14mm}

\author{
Lance Young$^1$, Contessa Wilburn$^1$, Carrie Houston$^1$, Blaine Keyton$^1$, Marwan Elawady$^1$,\\ Mohamed Shaban$^{1,2}$, and Muhammad Ismail$^1$
\vspace{1mm}
\\
{$^1$Cybersecurity Education, Research, and Outreach Center (CEROC) and Department of Computer Science,
}\\
{Tennessee Tech University, Cookeville, TN, USA.}\\
{$^2$Department of Mathematics, Faculty of Education, Alexandria University, Egypt} \\

Emails: \{layoung43, crwilburn42, cihouston42, bckeyton42, mmelawady42, mshaban, mismail\}@tntech.edu

\thanks{This work was supported by NSF Award \# 2322594.}
}
}
\maketitle

\begin{abstract}
    Quantum communication is a key enabler of next-generation networks, leveraging quantum entanglement to enable a new class of information exchange. While prior work has focused on the theoretical analysis of communication protocols, their exposure to physical layer side channel analysis remains largely unexplored. In classical systems, side channel analysis has been shown to reveal sensitive information without accessing the underlying data, raising the question of whether similar risks exist in quantum networks. In this work, we investigate whether different quantum communication protocols exhibit distinguishable signatures that can be inferred through passive side channel observations. We consider a threat model in which an observer accesses only a fraction of the optical signal without directly measuring the encoded quantum states. Under this setting, we experimentally examine four representative protocols, namely entanglement distribution, quantum gate sequences, heralded quantum key distribution, and quantum identity authentication, realized on a polarization entangled photon link. Observable physical layer features, including single photon detection statistics and optical power measurements, are collected and used to construct protocol fingerprints. We develop a data-driven framework for protocol identification based on these observations. Our results show that protocol identity can be inferred with accuracy reaching up to $96\%$ under $30{:}70$ sampling configuration/optical tapping, while remaining distinguishable at $10{:}90$ with accuracy ranging from $70{-}89\%$. Importantly, Bell inequality measurements confirm that the sampling/tapping process preserves entanglement, validating the non-destructive nature of the observation model. These findings demonstrate that side channel analysis can expose protocol-level information without disrupting quantum correlations, introducing new security considerations.
\end{abstract}


\section{Introduction}
\IEEEPARstart{Q}{uantum} communication networks are emerging as a key enabler of several applications, including distributed quantum computing, distributed quantum sensing, and secure communication services such as quantum key distribution (QKD) and quantum identity authentication (QIA)\cite{coupel2025securityvulnerabilitiesquantumcloud}. Unlike classical networks, these systems provide strong theoretical security guarantees rooted in the principles of quantum mechanics. As a result, significant research efforts have focused on designing protocols that ensure confidentiality and integrity at the quantum level. However, as quantum networks transition from theoretical models to practical implementations, new challenges arise from the physical realization of these systems. One important yet underexplored challenge is the potential leakage of information through physical layer side channels. In classical systems, side channel analysis has been widely used to infer sensitive information by observing indirect signals such as timing, power consumption, or electromagnetic emissions, without accessing the underlying data \cite{fingerprint_rubenstein,app9091881}. In contrast, quantum systems impose fundamental constraints on direct measurement, as measurement collapses the quantum state. While quantum communication protocols are designed to protect the quantum states carrying information, they rely on physical components such as photon sources, detectors, and optical channels \cite{gao2025spectralchannelsquantumkey}. These components may exhibit observable patterns that vary depending on the protocol being executed, raising the question of whether an external observer, with only passive access to physical layer signals, can distinguish which quantum communication protocol is currently running without directly measuring the quantum states. Answering this question is important because protocol fingerprinting can reveal network activity patterns, enable traffic analysis and usage profiling, and potentially facilitate targeted attacks on specific quantum services \cite{Rethinasamy_2023,Pantoja2024}. At the same time, such capabilities may also support non-intrusive monitoring and diagnostics for quantum networks, highlighting both security risks and system-level opportunities.

\subsection{Related Works}
\vspace{-2mm}
Existing work on side channel security in quantum systems has largely focused on specific implementations of QKD and quantum computing platforms rather than protocol-level inference in quantum networks. The work in \cite{9620820} investigates side channel vulnerabilities in the parity computation phase of the key reconciliation process in QKD, showing that intermediate classical operations may leak sensitive information. Similarly, the work in \cite{8539703} demonstrates that even a single trace can be sufficient to extract information from QKD implementations. Experimental leakage at the physical layer has also been observed in \cite{9531968}, where imperfections in BB84 sources lead to distinguishable emissions, and in \cite{PhysRevApplied.20.054040}, which uses deep learning to exploit radio frequency side channels from QKD systems. These efforts highlight that practical QKD implementations are susceptible to side channel leakage. However, these works are largely focused on specific protocols and assume prior knowledge of the protocol under attack.

Beyond QKD, the work in \cite{Erata_2024} shows that power side channels from quantum computer controllers can be used to reconstruct executed quantum circuits, revealing computational behavior through indirect observations. Similarly, \cite{10.1145/3716368.3735264} and \cite{11311002} explore timing side channels in cloud-based and simulated quantum computing environments, showing that execution patterns and system characteristics can be inferred without the measurement of quantum states. Additional studies on controller-level power leakage, such as \cite{10.1145/3576915.3623118}, further highlight the susceptibility of quantum hardware to physical side channel analysis. While these works demonstrate the feasibility of inferring computation or system behavior from indirect measurements, they primarily target quantum computing platforms rather than quantum communication links.

Despite these advances, the problem of distinguishing which quantum communication protocol is currently running in the network based solely on passive physical layer observations remains largely unexplored. In practical settings, an adversary may not know which protocol is active in the network before launching a targeted attack, making protocol identification a critical first step. 

\subsection{Contributions}
To address this gap, we experimentally investigate protocol fingerprinting across multiple quantum communication protocols. The main contributions of this paper are as follows:
\begin{itemize}
\item We introduce the problem of protocol fingerprinting in quantum communication networks and formalize a passive observation model where an external observer monitors the communication link without attempting to measure the encoded quantum information. We experimentally validate this model by implementing multiple quantum communication protocols, including entanglement distribution, quantum gate sequences, heralded QKD, and QIA, on a polarization-entangled photon testbed.
\item We develop a data-driven framework that leverages observable physical layer features, including single photon detection statistics and optical power measurements, and demonstrate that protocols can be distinguished with accuracy up to $96\%$ under a $30{:}70$ sampling configuration, while remaining distinguishable at $10{:}90$ with accuracy between approximately $70\%$ and $89\%$.
\item We experimentally show that passive sampling/tapping preserves entanglement while impacting entanglement rate, revealing a tradeoff between observability and rate reduction, and we provide interpretability analysis using Shapely additive explanations (SHAP) \cite{kraev2024shapselectlightweightfeatureselection} to show that timing features dominate classification decisions with complementary contributions from power-based measurements.
\end{itemize}

The remainder of this paper is organized as follows. Section~\ref{sec:system} defines the problem of protocol fingerprinting in quantum communication networks and presents the threat model and experimental testbed. Section \ref{sec:data} describes the implemented quantum communication protocols, along with the data collection process and feature engineering. Section \ref{sec:model} details the machine learning framework and methodology. Section \ref{sec:results} presents the experimental results, including entanglement validation, signal impact analysis, classification performance, and interpretability. Finally, Section \ref{sec:conclusion} concludes the paper and outlines future directions.

\section{Problem Definition, Threat Model, and Experimental Testbed} 
\label{sec:system}
This section formalizes the problem definition, introduction of the threat model, and presentation of the experimental testbed used in this study.
\subsection{Problem Definition}
We consider a quantum communication system in which multiple protocols are executed. These protocols include entanglement distribution \cite{PhysRevA.60.R773}, quantum gate sequences \cite{PhysRevA.105.012614}, heralded QKD \cite{PhysRevA.93.012331}, and QIA \cite{10757654}. Although these protocols differ in functionality, they rely on the same underlying optical components, including photon sources, detectors, and free-space optical channels. Let $\mathcal{P} = \{P_1, P_2, \dots, P_K\}$ denote the set of quantum communication protocols under consideration. Each protocol $P_k \in \mathcal{P}$ induces a distinct pattern of physical layer behavior due to differences in state preparation, measurement configurations, and temporal structure. An external observer passively monitors the system and collects a set of observable signals $\mathbf{x} \in \mathcal{X}$, where $\mathbf{x}$ represents features derived from physical layer measurements such as photon detection statistics and optical power levels. These observations are obtained without direct measurement of the underlying quantum states. The problem addressed in this work is to infer the protocol identity from the observed signals. Formally, this can be expressed as a classification problem
\begin{equation}
f: \mathcal{X} \rightarrow \mathcal{P},
\end{equation}
where $f(\cdot)$ maps observed physical layer features to the corresponding protocol label. This formulation captures a realistic setting in which only indirect observations are available, and the objective is to distinguish between protocols based solely on their physical layer signatures.

\subsection{Threat Model}
We consider an external observer that passively monitors the quantum communication link. The observer does not measure or otherwise interact with the quantum states directly, as such interaction would disturb the system and reveal the presence of an attack. Instead, the observer samples a fraction of the optical signal via optical tapping, enabling indirect observation while minimizing impact on the communication link. The observer is considered to have access only to physical layer measurements, including photon detection events and optical power readings. No prior knowledge of the transmitted quantum states, encoding schemes, protocol family or type is assumed. The observer aims to infer the protocol currently being executed by analyzing patterns in the observed signals. This setting reflects a passive adversarial scenario and also captures potential use cases for non-intrusive monitoring and diagnostics.

\subsection{Experimental Testbed}
\label{sec:testbed}
\begin{figure*}
    \centering
    \includegraphics[width=\linewidth]{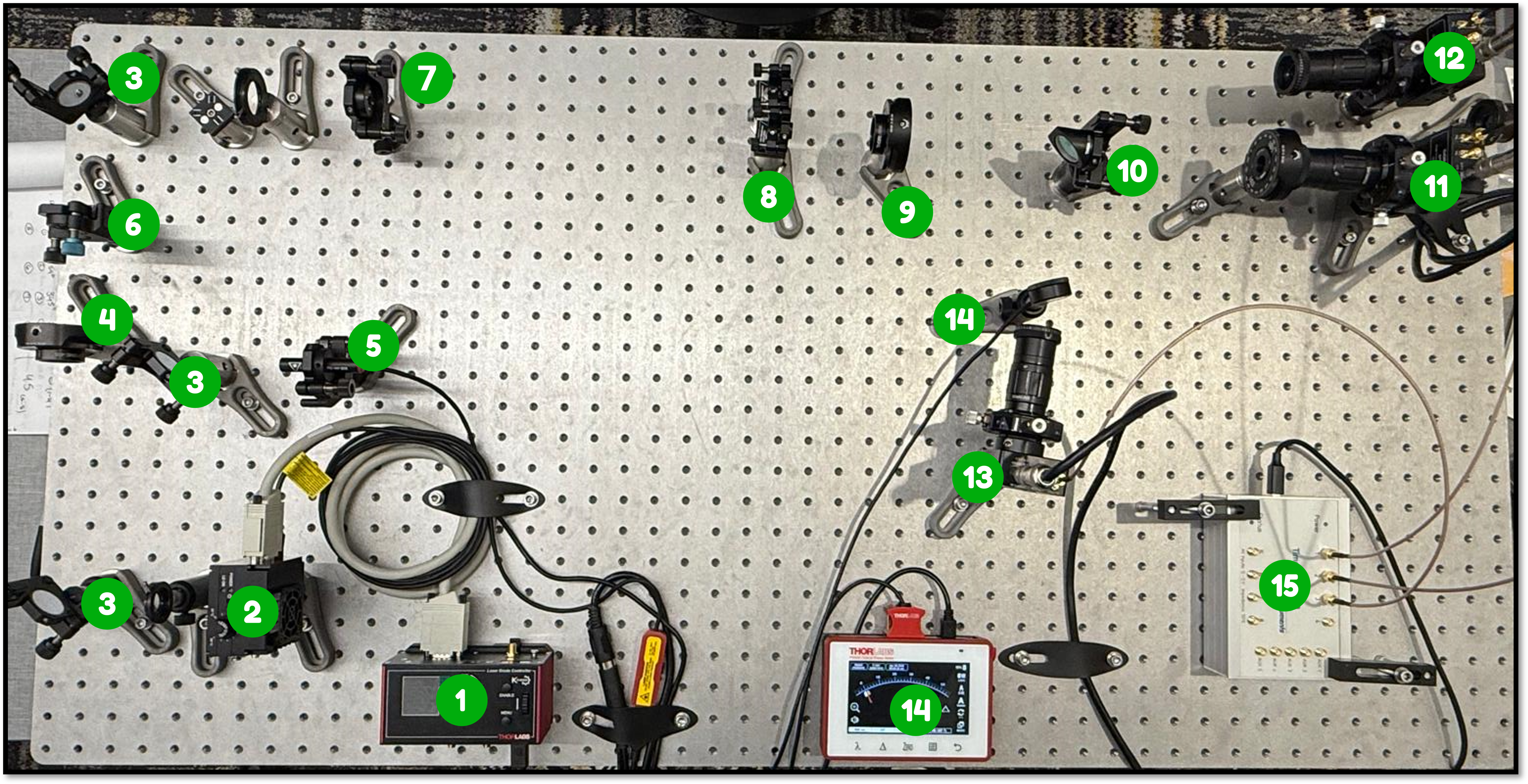}
    \caption{Quantum communication testbed used for protocol fingerprinting via physical layer observations. The setup includes (1) laser driver, (2) $405$ nm pump laser source, (3) kinematic mirrors for beam alignment, (4) $405$ nm half-wave plate, (5) alignment laser, (6) temporal compensation crystal, (7) BBO nonlinear crystal for entangled photon generation, (8) spatial compensation crystals, (9) $808$ nm half-wave plate for state preparation, (10) beam splitter for passive sampling, (11) single-photon avalanche photodetectors (signal photon receiver), (12) herald photon detector, (13) passive observer detector, (14) passive observer power meter, and (15) time-tagger.}
    \label{fig:testbed}
\end{figure*}

To facilitate this study, we utilize a polarization-entangled photon testbed that enables the implementation of multiple quantum communication protocols, as illustrated in Fig.~\ref{fig:testbed}.
The system is driven by a continuous-wave pump laser operating at $405$ nm, which is directed through a half-wave plate to control the polarization of the pump beam. The beam then passes through a temporal compensation crystal to correct phase mismatch between polarization components. The pump is subsequently directed to a crossed beta-barium borate (BBO) crystal configuration, where spontaneous parametric down-conversion (SPDC) occurs, generating pairs of polarization-entangled photons. By properly adjusting the phase, the generated state can be expressed as
\begin{equation}
|\psi\rangle = \frac{1}{\sqrt{2}} \left(|HH\rangle + e^{i\phi} |VV\rangle \right),
\end{equation}
where $\phi$ is the relative phase adjusted by the compensation stage, enabling the generation of a maximally entangled Bell state. The quality of entanglement was verified using a standard Bell (CHSH) test, confirming non-classical correlations between the photon pairs. The generated photon pairs are separated into two optical paths and detected using single-photon avalanche photodiodes. The detector outputs are connected to a time-tagger, which records photon arrival times with picosecond resolution, allowing for precise extraction of photon detection statistics and coincidence measurements. To emulate the external observer described in the threat model, a beam splitter is introduced along one of the down-converted photon paths to passively sample a fraction of the optical signal. The tap is positioned after the state preparation stage (i.e., the 808 nm half-wave plate) and before the measurement components, such as the polarizer and detector.  The sampled output is directed to an observation unit consisting of either a single-photon detector or a power meter. The observer is limited to measuring aggregate physical layer quantities, such as photon detection statistics and optical power, without performing direct measurements on the quantum states or introducing any active interference. Different quantum communication protocols are implemented on the same testbed by configuring polarization control and measurement components. In particular, an $808$ nm half-wave plate is used to prepare different quantum states, while protocol-specific measurement configurations are realized through polarizers or detector roles, such as heralded detection. These variations induce distinct temporal and statistical patterns in the observed signals. Detailed descriptions of individual protocol implementations and data collection procedures are provided in Section \ref{sec:data}. This shared hardware setup ensures that observable differences arise from protocol behavior rather than changes in the underlying physical infrastructure.

\section{Protocol Fingerprinting}
\label{sec:data}
This section presents the communication protocols under study, data collection, and feature engineering methods.
\subsection{Quantum Communication Protocols}
In this work, we consider multiple quantum communication protocols executed on the same physical testbed, including entanglement distribution, quantum gate sequences, heralded QKD, and QIA. Although these protocols share the same hardware infrastructure, they differ in state preparation, measurement configurations, and temporal structure, which leads to distinguishable physical layer signatures. The protocols considered in this study are summarized next:
\begin{itemize}
    \item \textit{Entanglement Distribution \cite{PhysRevA.60.R773}:}
    In this protocol, polarization-entangled photon pairs are generated and distributed between two parties without additional encoding. Measurements are performed to observe correlations between photon pairs, resulting in relatively stable photon detection statistics and temporal behavior. This protocol serves as a baseline, as it introduces minimal modulation beyond the entanglement source.
    
    \item \textit{Quantum Gate Sequences \cite{PhysRevA.105.012614}:}
    This protocol emulates simple quantum processing behavior by applying sequences of single-qubit transformations through controlled polarization rotations. A half-wave plate operating at $808$ nm acts as a programmable processing stage, where the polarization state of the signal photon is modified prior to measurement. Different operation patterns are constructed by varying the orientation of the half-wave plate over consecutive time intervals. In this work, we implement two representative sequences, $\{0^\circ, 45^\circ, 0^\circ, 45^\circ\}$ and $\{0^\circ, 22.5^\circ, 45^\circ, 67.5^\circ\}$, corresponding to alternating and multi-transition polarization transformations, respectively. These rotations correspond to standard single-qubit operations, where $\theta = 45^\circ$ implements a Pauli-X transformation and $\theta = 22.5^\circ$ implements a Hadamard transformation. The signal photon propagates from the processing stage to the measurement point, where it is detected using a polarization analyzer and single-photon detector. An external observer passively monitors the communication channel between the processing stage and the measurement endpoint through a tap, as described in Section~\ref{sec:testbed}. The second photon of the entangled pair is used as a herald to indicate the presence of the corresponding signal photon.
     
    \item \textit{Heralded QKD \cite{PhysRevA.93.012331}:} In the heralded QKD protocol, one photon of the entangled pair is used as a herald to indicate the presence of the corresponding signal photon. The signal photon is prepared in one of four quantum states, $\ket{H}$, $\ket{V}$, $\ket{D}$, and $\ket{A}$, corresponding to horizontal, vertical, diagonal, and anti-diagonal polarizations. State preparation is achieved using an $808$ nm half-wave plate, while measurement is performed using a linear polarizer that selects the measurement basis. The preparation states and measurement bases are chosen randomly, introducing variability in photon detection statistics and correlation patterns. This variability leads to characteristic physical layer signatures associated with the protocol.

    \item \textit{QIA \cite{10757654}:} This protocol is implemented in a heralded manner, where one photon is used as a herald to indicate the presence of the corresponding signal photon. The protocol consists of two operational phases, namely, data transmission and authentication. Specifically, it introduces periodic authentication rounds interleaved with quantum state transmission, where the frequency of authentication is determined by a shared secret key. During the data transmission phase, the signal photon is prepared in either $\ket{H}$ or $\ket{V}$, selected randomly to encode the payload. During the authentication phase, the transmitted states are chosen from the set $\{\ket{H}, \ket{V}, \ket{D}$, and $\ket{A}\}$ according to the shared secret key, following the encoding defined in \cite{10757654}. These alternating phases introduce structured, key-dependent temporal patterns in the transmitted signals, resulting in distinctive statistical and temporal characteristics observable at the physical layer.
\end{itemize}

\subsection{Data Collection}
\vspace{-1mm}
Data is collected from the experimental testbed by passively observing the optical signal through a beam splitter, as described in Section \ref{sec:system}. Two sampling/tapping configurations are considered to emulate different levels of observability, where only a fraction of the optical signal is available to the external observer. The sampled signal is directed to an observation unit consisting of either a single-photon detector or a power meter. Photon detection events are recorded using a time tagger, which provides high-resolution timestamps of photon arrivals. From these measurements, detection counts and timing characteristics of photon arrivals are obtained. Optical power measurements are collected using a power meter to capture variations in signal intensity over time. These measurements complement the photon detection data by providing an additional view of the system behavior at the physical layer. For each protocol, data is collected over multiple runs under consistent experimental conditions to ensure comparability. The collected data is segmented into fixed-duration intervals, where each segment corresponds to a specific protocol execution instance. During dataset construction, segments corresponding to different protocol executions are randomly arranged to prevent bias due to fixed temporal ordering. From the collected measurements, statistical features are extracted to characterize timing and intensity variations. The details of feature construction and preprocessing are described in Sections \ref{sec:features} and \ref{sec:preprocessing}.


\subsection{Feature Engineering}
\label{sec:features}
To enable protocol classification, we construct feature representations from the observable physical layer measurements collected by the external observer. These features are derived exclusively from the passive observer channel and capture both timing and intensity-based characteristics of the signal as follows:
\begin{itemize}
\item \textit{Time Tagger Features:} From the photon arrival timestamps, we extract the photon count rate, defined as the number of detected photons per unit time. In addition, we compute interarrival time (IAT) statistics, including the mean and standard deviation of the time intervals between consecutive photon detections. To capture relative variability, we also compute the coefficient of variation (CV) of the interarrival times, which reflects the degree of burstiness or irregularity in photon arrivals.
\item \textit{Power Meter Features:} From the sampled optical signal, we extract irradiance measurements and compute statistical features including the mean, standard deviation, and CV. These features capture fluctuations in optical intensity over time. In addition, raw power measurements are recorded in both milliwatts and decibel units (dBm), providing complementary representations of signal strength.
\end{itemize}
The selected features are designed to capture both the magnitude and variability of the observed signals. Time-based features reflect photon arrival dynamics, while power-based features capture intensity fluctuations. Together, these features provide a comprehensive characterization of physical layer behavior, enabling differentiation between protocols based on their temporal and statistical signatures.

\subsection{Data Preprocessing}
\label{sec:preprocessing}
To improve robustness and mitigate the impact of experimental variability, all features are normalized using a local min-max scaling approach. Data is collected across multiple runs, where environmental factors such as temperature fluctuations and minor alignment variations may introduce shifts in the observed measurements. Normalization ensures that the learning process focuses on relative patterns in the data rather than absolute magnitudes that may vary across experiments. Outliers are mitigated using percentile-based clipping to improve robustness to measurement noise and transient fluctuations arising from detector effects (e.g., dark counts and afterpulsing) and optical background variations. The resulting feature values, $x$, are then normalized using min--max scaling, which maps each feature to the range $[0,1]$ according to
\begin{equation}
x_{\text{norm}} = \frac{x - x_{\min}}{x_{\max} - x_{\min}}.
\end{equation}
The dataset is constructed from a balanced subset of $7,200$ samples. To capture temporal dependencies, the data is segmented into fixed-length sequences of five consecutive samples, resulting in $1,440$ sequence-based samples. The dataset is then split into $80\%$ for training and $20\%$ for testing, corresponding to $1,152$ training samples and $288$ testing samples. Non-overlapping sequences are used so that each sample represents a distinct temporal window, avoiding redundancy and reducing the risk of information leakage between training and testing sets.



\section{Machine Learning-Based Fingerprinting}
\label{sec:model}
This section presents the considered machine learning models for protocol fingerprinting and the conducted interpretability analysis.
\subsection{Model Training and Evaluation}
To infer protocol identity from physical layer observations, the feature representations described in Section \ref{sec:data} are used as inputs to train sequence-based machine learning models. Given the temporal nature of the collected data, we employ recurrent neural network (RNN) architectures, including Long Short-Term Memory (LSTM), Gated Recurrent Unit (GRU), and their bidirectional and stacked variants. These models are well-suited for capturing temporal dependencies and sequential patterns in the observed signals. Hyperparameters for all models are tuned empirically to achieve stable convergence and strong classification performance. The best-performing model in most cases, the Bi-Stacked LSTM, consists of two recurrent layers with $64$ and $32$ hidden units, respectively, with sequence return enabled in the first layer. A dropout rate of $0.3$ is applied between layers for regularization. The resulting representation is passed through a dense layer with $32$ units and a hyperbolic tangent (tanh) activation before the output layer. The model is trained using the Adam optimizer with a learning rate of $10^{-4}$ and categorical cross-entropy loss. The dataset is divided into training and testing sets using an $80/20$ split, ensuring that evaluation is performed on unseen data. Model performance is evaluated using standard classification metrics, including accuracy and F1-score.
\subsection{Feature Importance and Interpretability}
To understand the contribution of individual features to protocol classification, we employ Shapley Additive explanations (SHAP) \cite{10.5555/3295222.3295230, kraev2024shapselectlightweightfeatureselection}. SHAP provides a model-agnostic framework for quantifying the impact of each feature on the prediction outcome. By analyzing SHAP values across different protocols, we identify which physical layer characteristics contribute most to protocol distinguishability. In particular, timing features derived from photon detection events capture variations in arrival dynamics, while power-based features reflect intensity fluctuations. The combination of these features enables the model to distinguish between protocols based on their underlying physical layer signatures. This interpretability analysis provides insight into the physical mechanisms driving classification performance and validates the effectiveness of the selected feature set.



\section{Experimental Results and Analysis}
\label{sec:results}
We evaluate the effectiveness of protocol fingerprinting under different sampling/tapping ratios and feature configurations. The classification results are summarized in Table~\ref{tab:results}. The analysis of these results is detailed in the following subsections.

\begin{table*}[!ht]
\renewcommand{\arraystretch}{1.4} 
\centering
\caption{Protocol classification accuracy (\%) and F1-Score under different sampling ratios and feature sets.}
\label{tab:results}
\footnotesize 
\setlength{\tabcolsep}{3pt} 
\begin{tabular}{|l|cc|cc|cc|cc|cc|cc|}
\hline
\multirow{3}{*}{\textbf{Model}} & \multicolumn{4}{c|}{\textbf{Time Tagger}} & \multicolumn{4}{c|}{\textbf{Power Meter}} & \multicolumn{4}{c|}{\textbf{Combined}} \\ \cline{2-13} 
 & \multicolumn{2}{c|}{\textbf{10:90}} & \multicolumn{2}{c|}{\textbf{30:70}} & \multicolumn{2}{c|}{\textbf{10:90}} & \multicolumn{2}{c|}{\textbf{30:70}} & \multicolumn{2}{c|}{\textbf{10:90}} & \multicolumn{2}{c|}{\textbf{30:70}} \\ \cline{2-13} 
 & \textbf{Accuracy} & \textbf{F1-Score} & \textbf{Accuracy} & \textbf{F1-Score} & \textbf{Accuracy} & \textbf{F1-Score} & \textbf{Accuracy} & \textbf{F1-Score} & \textbf{Accuracy} & \textbf{F1-Score} & \textbf{Accuracy} & \textbf{F1-Score} \\ \hline
SimpleRNN & 71.88 & 0.72 & 79.17 & 0.78 & 68.06 & 0.67 & 82.64 & 0.82 & 69.79 & 0.70 & 89.93 & 0.90 \\ \hline
LSTM & 72.22 & 0.72 & 77.43 & 0.76 & 61.46 & 0.60 & 83.33 & 0.83 & 77.78 & 0.78 & 89.58 & 0.89 \\ \hline
Bi-LSTM & 79.51 & 0.79 & 87.50 & 0.87 & 67.71 & 0.67 & 85.07 & 0.85 & 86.81 & 0.87 & 92.71 & 0.93 \\ \hline
Bi-GRU & 79.17 & 0.79 & 90.28 & 0.90 & 68.75 & 0.68 & 86.81 & 0.87 & 84.38 & 0.84 & 93.06 & 0.93 \\ \hline
\textbf{Bi-Stacked LSTM} & \textbf{87.50} & \textbf{0.87} & \textbf{96.18} & \textbf{0.96} & \textbf{73.61} & \textbf{0.73} & \textbf{84.72} & \textbf{0.85} & \textbf{89.24} & \textbf{0.89} & \textbf{95.49} & \textbf{0.96} \\ \hline
\end{tabular}
\end{table*}

\subsection{Impact of Passive Sampling on Entanglement, Photon Count, and Coincidence Rates}
To evaluate whether passive optical sampling disrupts the underlying quantum correlations, we measure the CHSH Bell parameter (S-value) under different sampling configurations. Without tapping, the system achieves an S-value of $2.347 \pm 0.014$. With a $30{:}70$ split, the S-value is $2.324 \pm 0.016$, and with a $10{:}90$ split, it is $2.359 \pm 0.014$. In all cases, the measured S-values remain well above the classical bound of $2$, confirming that entanglement is preserved under the considered sampling ratios.

\begin{figure}[t]
\centering
\includegraphics[width=\linewidth]{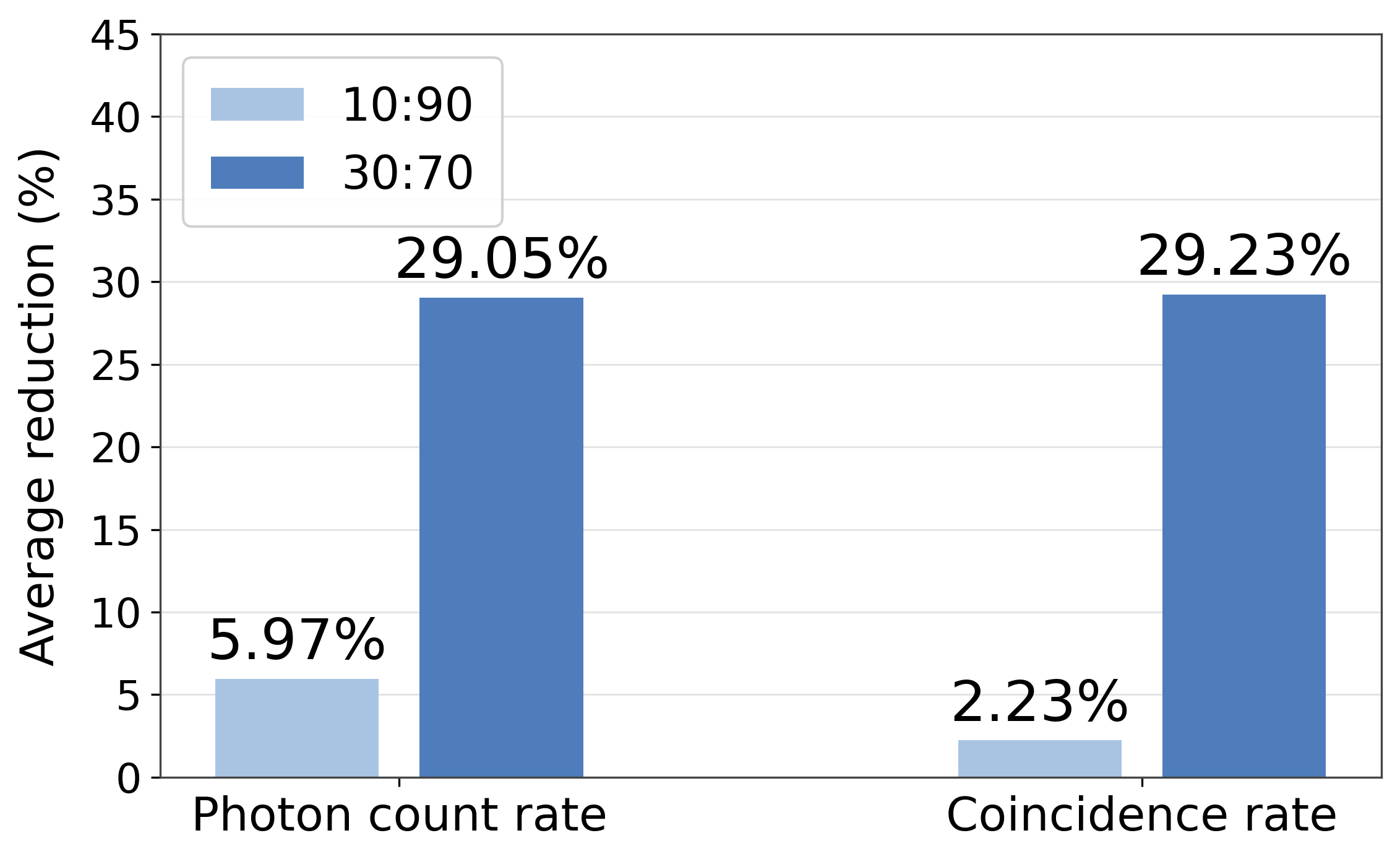}
\caption{Effect of passive sampling on photon count and coincidence rates. The $30{:}70$ configuration introduces significantly higher reductions in photon count rate and coincidence rate compared to the $10{:}90$ configuration.}
\label{fig:sampling_effect}
\end{figure}

In addition to entanglement validation, we quantify the impact on photon count and coincidence rates introduced by passive sampling. The $30{:}70$ split reduces the photon count rate by approximately $29.05\%$ and the coincidence rate by approximately $29.23\%$, while the $10{:}90$ split results in smaller reductions of approximately $5.97\%$ and $2.23\%$, respectively, as shown in Fig.~\ref{fig:sampling_effect}. These results show that increasing the sampling ratio leads to substantial reduction in photon count and coincidence rates, while entanglement remains preserved. Taken together, these findings validate the non-destructive nature of the proposed observation model while revealing a clear tradeoff between observability and entanglement rate. The $30{:}70$ configuration provides richer side channel information for protocol inference, as shown in Table~\ref{tab:results}, at the cost of reductions in the coincidence rate, whereas the $10{:}90$ configuration maintains significantly lower reduction in coincidence rate while still preserving reliable distinguishability.

\subsection{Impact of Sampling Ratio}
A clear trend across all models and feature sets is the impact of the sampling ratio on classification performance. Higher sampling ratios consistently yield improved accuracy, with the $30{:}70$ configuration achieving the best overall results. This behavior is expected, as increased access to the optical signal provides richer information about the underlying protocol behavior. However, higher sampling ratios also imply greater interaction with the communication channel, which may affect normal network operation. In contrast, the $10{:}90$ configuration represents a more constrained observation setting, where a small fraction of the signal is accessible. Despite the reduced observability, the results show that protocol identity remains distinguishable, with accuracy remaining relatively high across models. This indicates that meaningful protocol-level information can be inferred even under limited access to the signal. However, as the sampling ratio decreases, the available information becomes more limited, leading to increased uncertainty in protocol identification due to reduced visibility of temporal and statistical patterns.
\subsection{Impact of Feature Sets}
The choice of feature set plays a significant role in classification performance. Time tagger features consistently outperform power meter features when used independently, indicating that timing characteristics of photon arrivals provide stronger discriminative information. In contrast, power meter features alone achieve lower accuracy, suggesting that intensity-based measurements capture less protocol-specific variation.  When both feature sets are combined, classification performance improves across most models, achieving the highest overall accuracy. An exception is observed in the $10{:}90$ configuration for the SimpleRNN model, which exhibits reduced performance when handling multi-modal inputs compared to gated architectures such as LSTM and GRU. From an observation perspective, each feature type reflects a different level of access to the communication link. Time tagger measurements provide the most informative view, enabling higher classification accuracy, while power meter measurements remain effective, achieving accuracy in the range of $74\%$ to $85\%$. These results demonstrate that timing and intensity-based observations provide complementary information, enabling more reliable protocol fingerprinting.
\subsection{Model Performance}
The results also highlight the importance of model architecture in capturing protocol-specific patterns. Simpler models, such as standard RNNs and single-layer LSTMs, achieve moderate performance. In contrast, more advanced architectures, including bidirectional and stacked recurrent models, consistently achieve higher accuracy. In particular, the Bi-Stacked LSTM achieves the best performance across most configurations, reaching up to $96\%$ accuracy under the $30{:}70$ sampling ratio and $95\%$ when using the combined feature set. This indicates that deeper architectures with bidirectional processing are more effective at capturing complex temporal dependencies in the observed signals.

\subsection{Error Analysis via Confusion Matrices}
\begin{figure*}
\centering
\subfloat[]{\includegraphics[width= 2.3in]{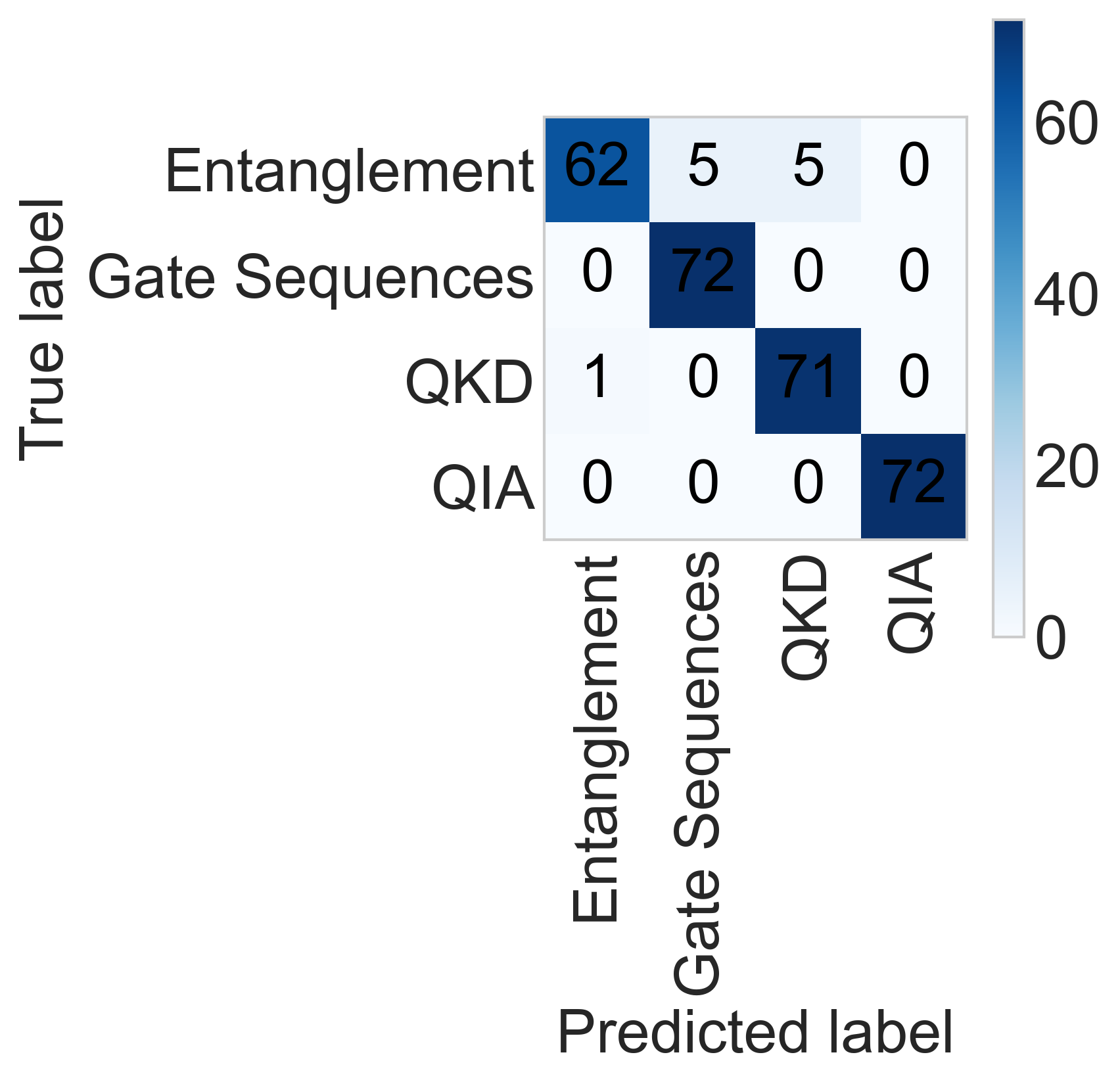}\label{fig:confusion_30_70_tagger}} \;\;
\subfloat[]{\includegraphics[width= 2.3in]{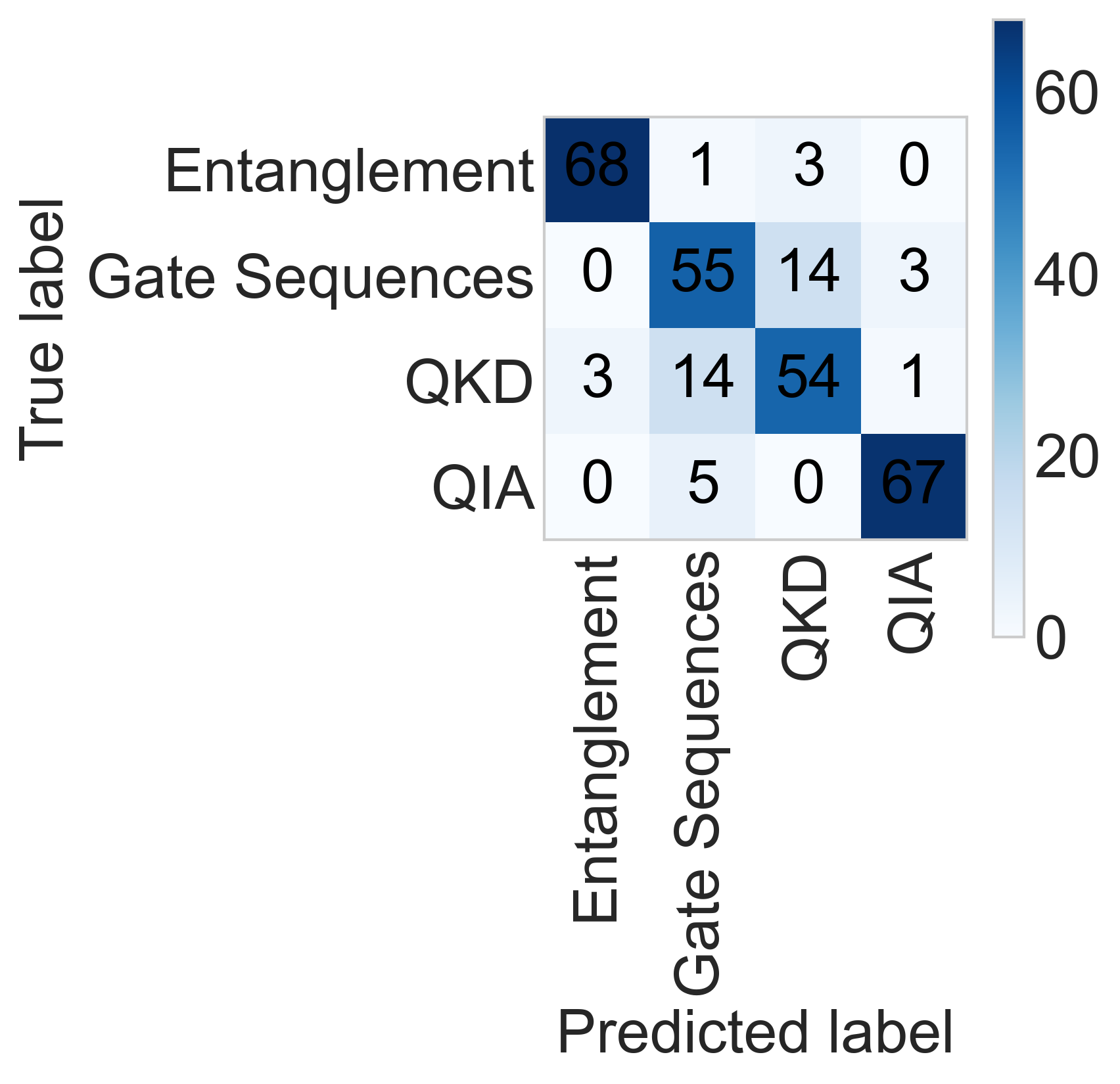}\label{fig:confusion_30_70_power}} \;\;
\subfloat[]{\includegraphics[width= 2.3in]{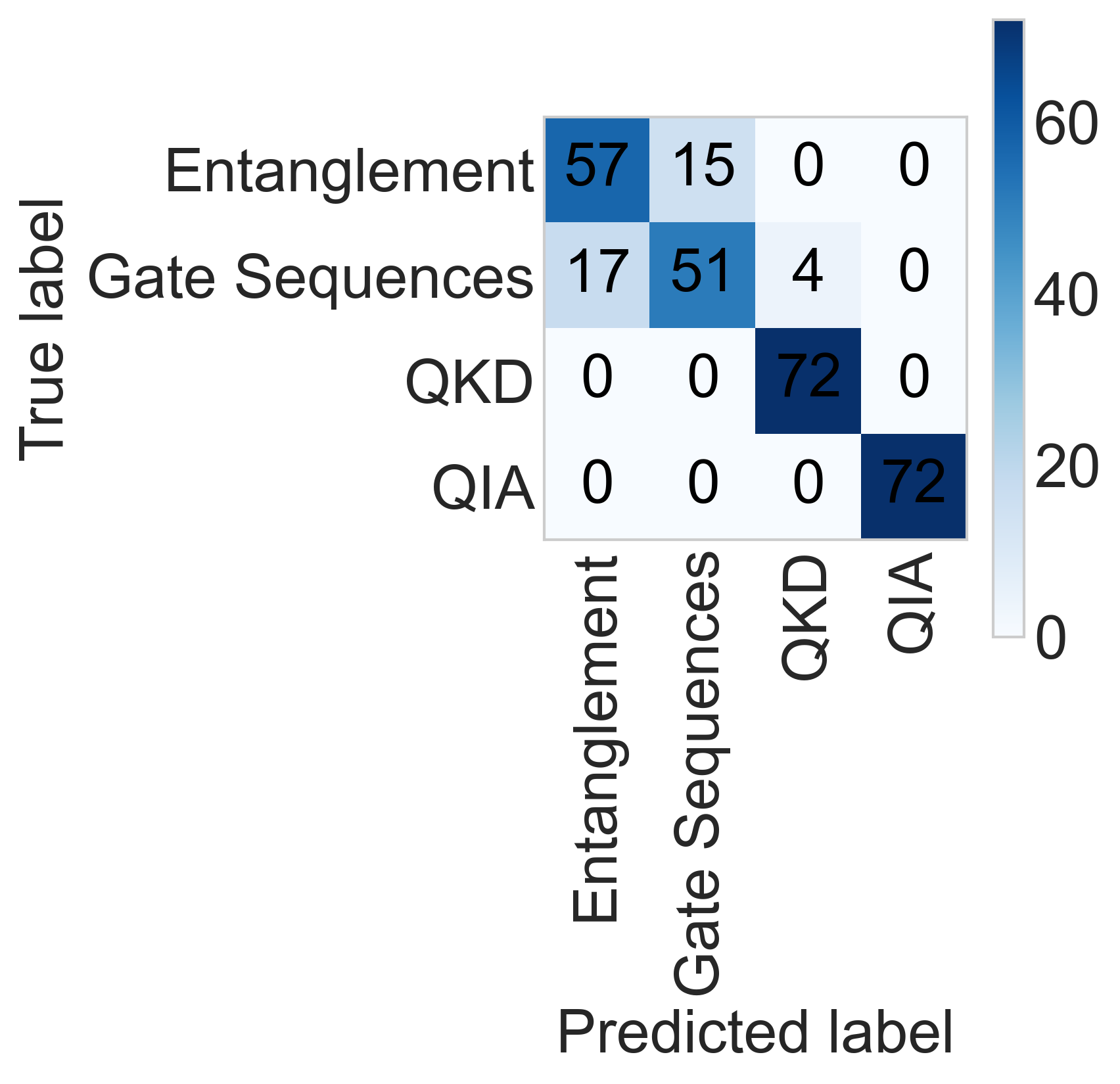}\label{fig:confusion_10_90_tagger}} \\
\subfloat[]{\includegraphics[width= 2.3in]
{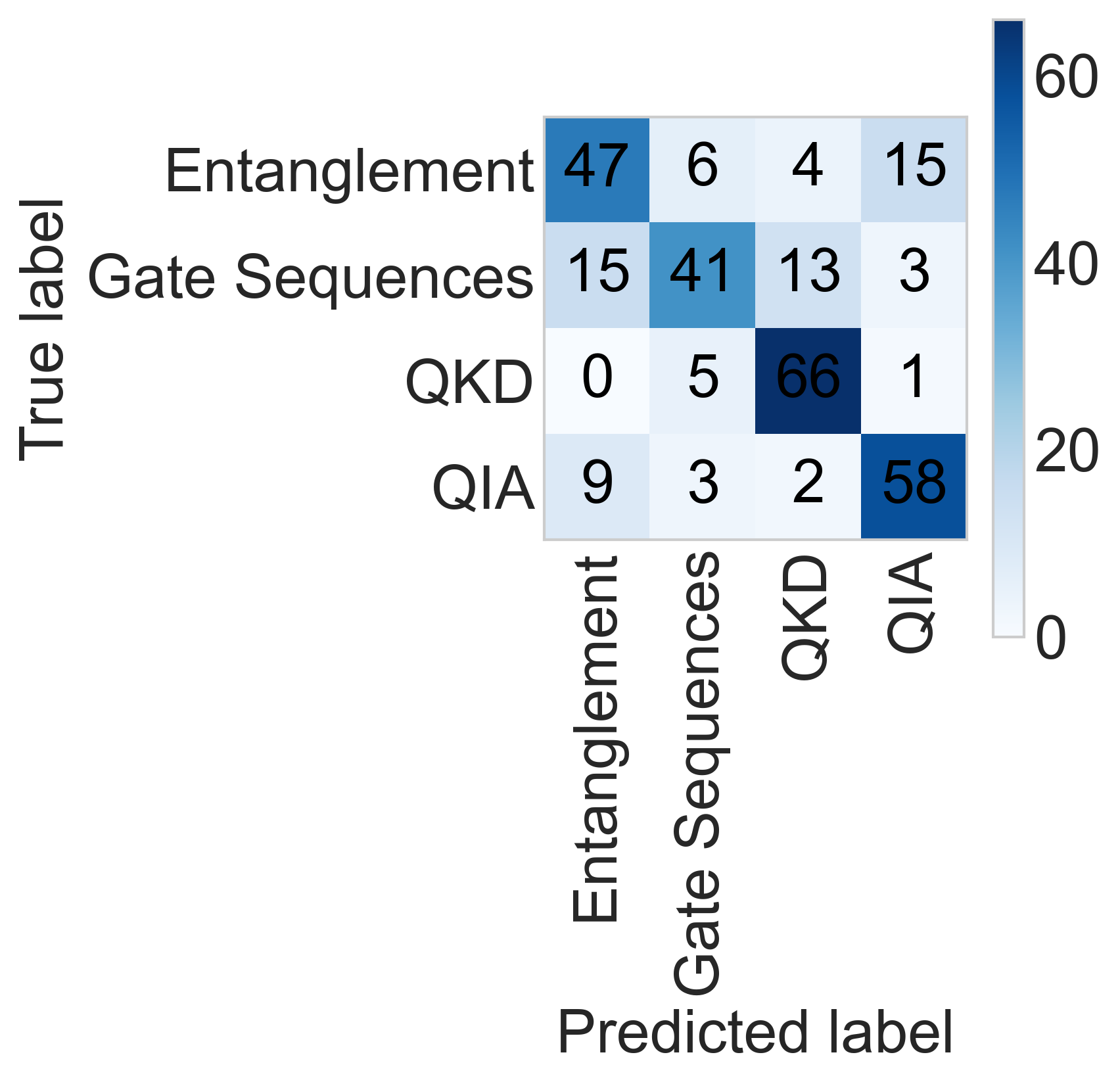}\label{fig:confusion_10_90_power}} \;\;
\subfloat[]{\includegraphics[width= 2.3in]
{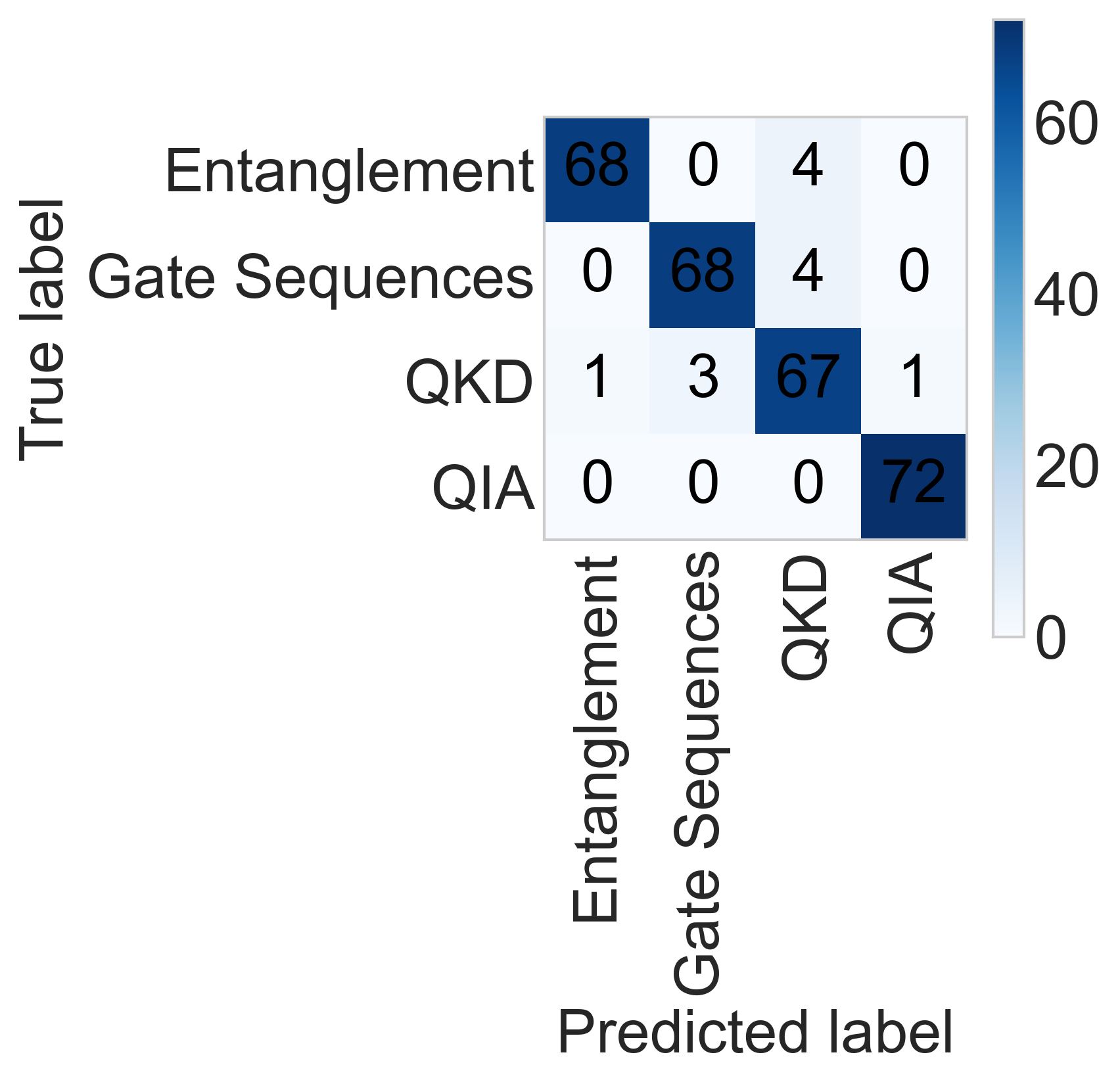}\label{fig:confusion_30_70_combined}} \;\;
\subfloat[]{\includegraphics[width= 2.3in]
{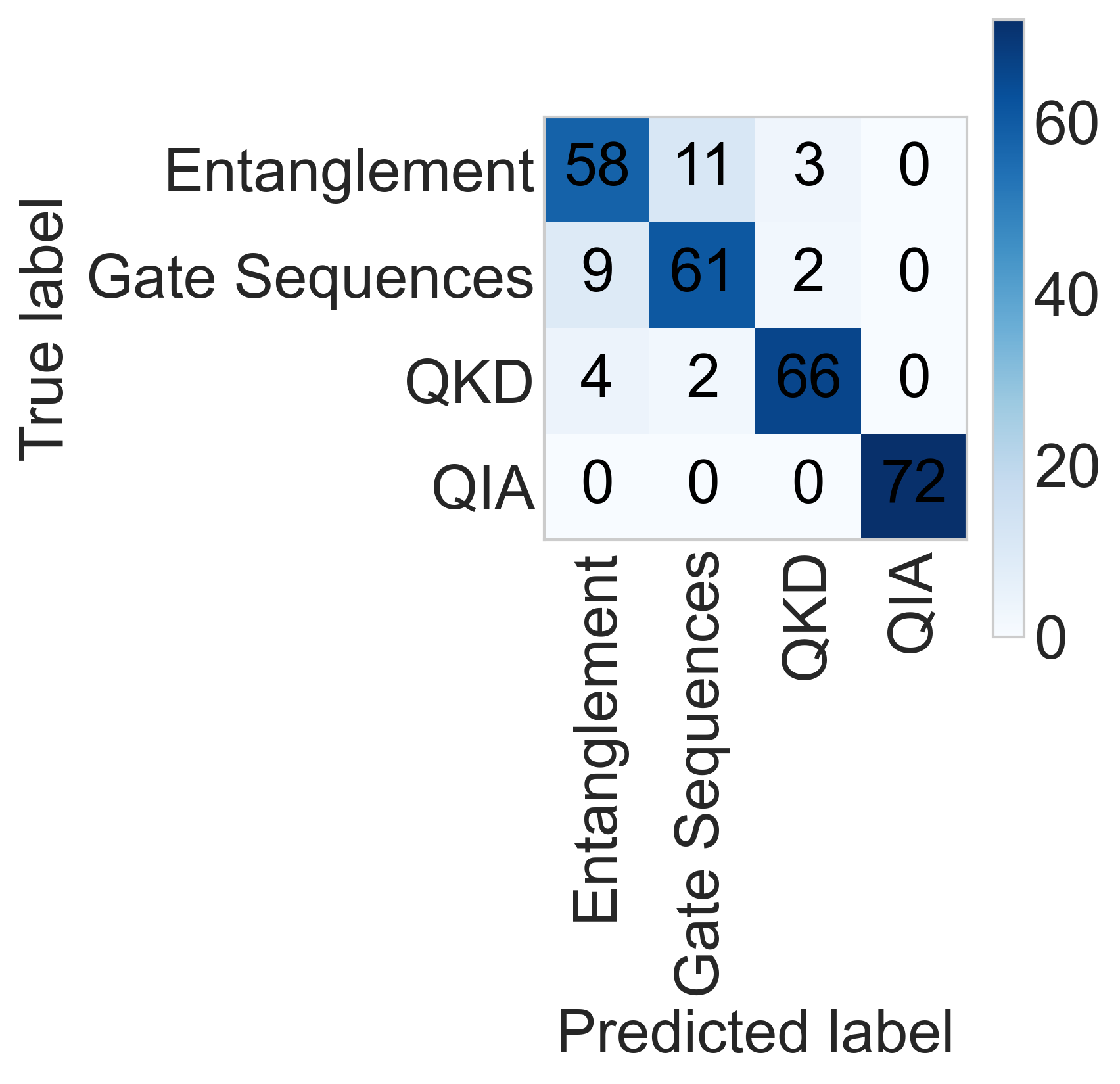}\label{fig:confusion_10_90_combined}}

\caption{Confusion matrices illustrating the impact of sampling ratio and feature type on protocol classification. (a) Time tagger features at $30{:}70$, (b) power meter features at $30{:}70$, (c) time tagger features at $10{:}90$, (d) power meter features at $10{:}90$, (e) combined features at $30{:}70$, and (f) combined features at $10{:}90$. Higher sampling ratios improve class separability, while time-resolved features provide stronger discriminative capability than intensity-based measurements. The combination of both feature types yields the highest overall performance.}
\label{fig:matrices}
\end{figure*}

To further understand the classification behavior, we analyze confusion matrices under different sampling ratios and feature configurations, as shown in Fig.~\ref{fig:matrices}. 
Since the Bi-Stack LSTM consistently achieves the highest classification performance, the confusion matrices presented correspond to this model. Comparing Figs.~\ref{fig:confusion_30_70_tagger} and \ref{fig:confusion_10_90_tagger}, which correspond to time tagger features under the $30{:}70$ and $10{:}90$ sampling configurations, respectively, highlights the impact of sampling ratio on class separability. At higher sampling levels, the confusion matrix exhibits a strong diagonal structure, indicating accurate classification with minimal confusion between protocols. In contrast, under the $10{:}90$ configuration, structured misclassifications become more apparent, suggesting that reduced observability limits the ability to fully capture distinguishing temporal patterns. The impact of feature type is illustrated by comparing Figs.~\ref{fig:confusion_10_90_tagger} and \ref{fig:confusion_10_90_power}, which correspond to time tagger and power meter features under the same $10{:}90$ sampling configuration. Time tagger features produce a more pronounced diagonal structure, indicating stronger class separation, while power meter features exhibit more dispersed misclassifications. This demonstrates that temporal characteristics of photon arrivals provide more discriminative information than intensity-based measurements. Finally, Figs.~\ref{fig:confusion_30_70_combined} and \ref{fig:confusion_10_90_combined} show that combining both feature types improves classification performance across sampling configurations. The combined feature set reduces misclassification and yields more concentrated diagonal patterns. Overall, the confusion matrix analysis reveals that classification errors are not random but arise from similarities in the physical layer signatures of certain protocols. These structured errors diminish with increased sampling and the use of complementary feature types, supporting the effectiveness of the proposed fingerprinting approach. Overall, the confusion matrix analysis reveals that classification errors are not random but arise from similarities in the physical layer signatures of certain protocols. The reduction of these structured errors with higher sampling ratios and the use of complementary feature sets further supports the effectiveness of the proposed fingerprinting approach.

\subsection{Model Interpretability via SHAP Analysis}
\begin{figure*}[]
\centering
\subfloat[]{\includegraphics[width=3.6in]{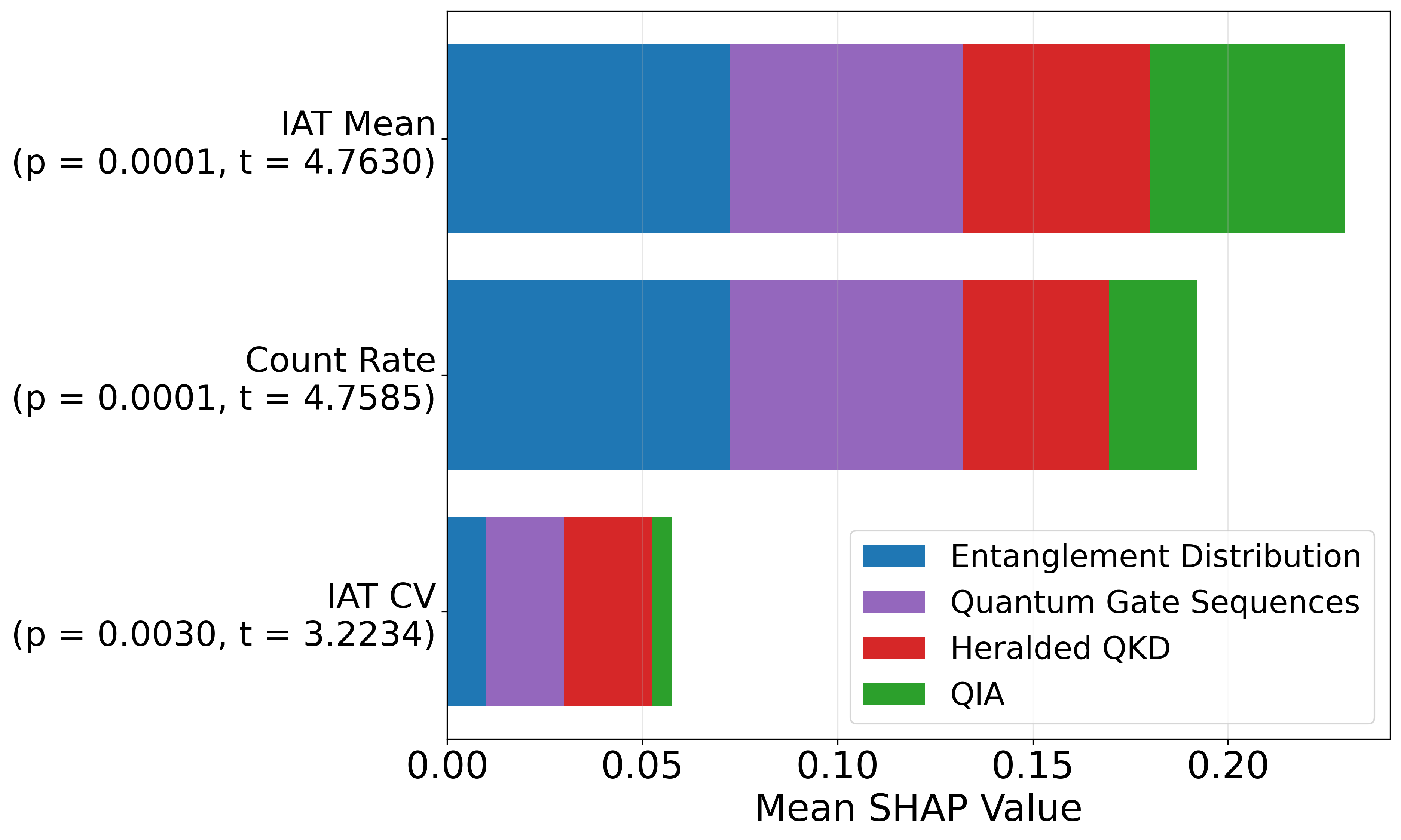}\label{shap_1090_tagger}}
\subfloat[]{\includegraphics[width=3.6in]{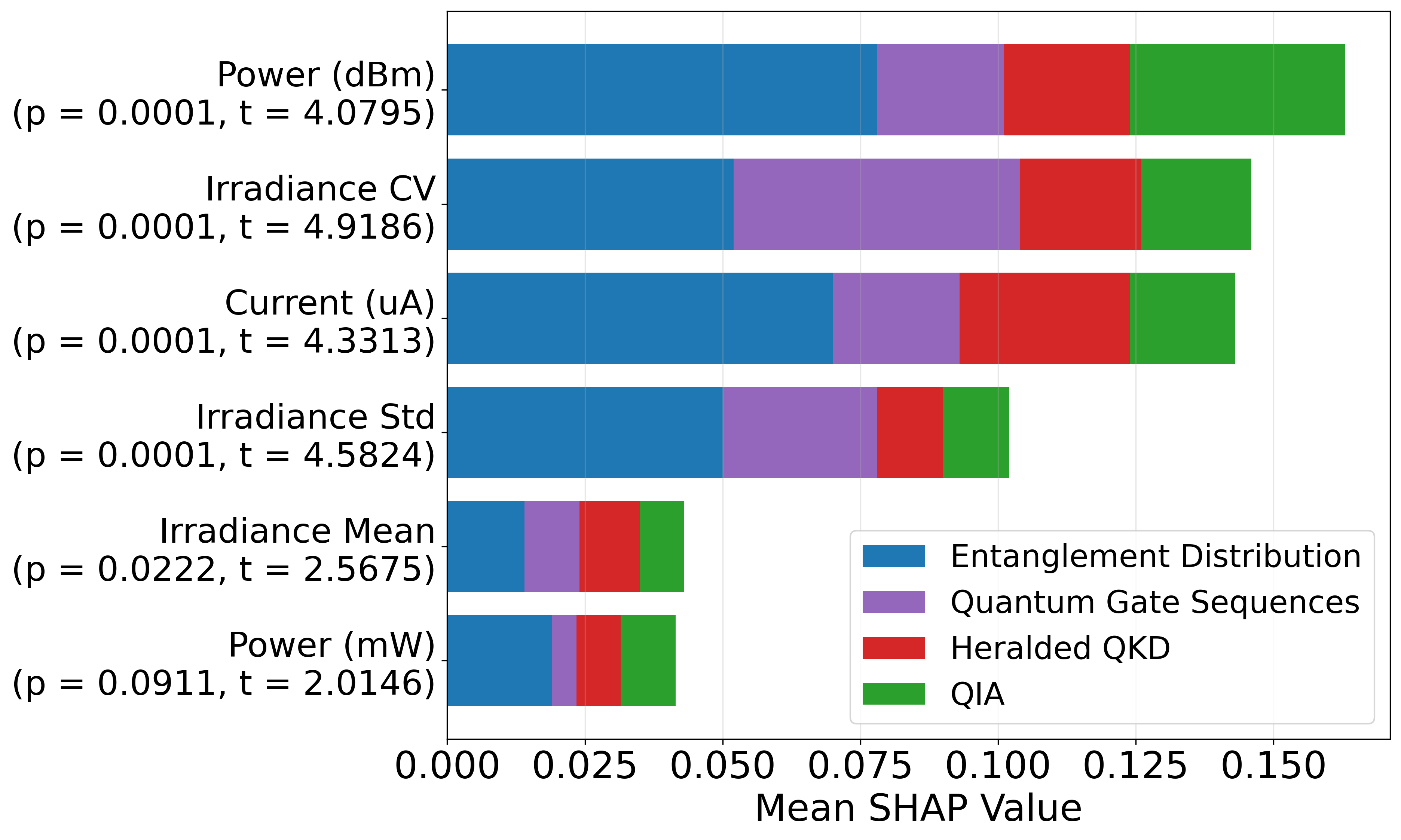}\label{shap_1090_power}} \\
\subfloat[]{\includegraphics[width=3.6in]{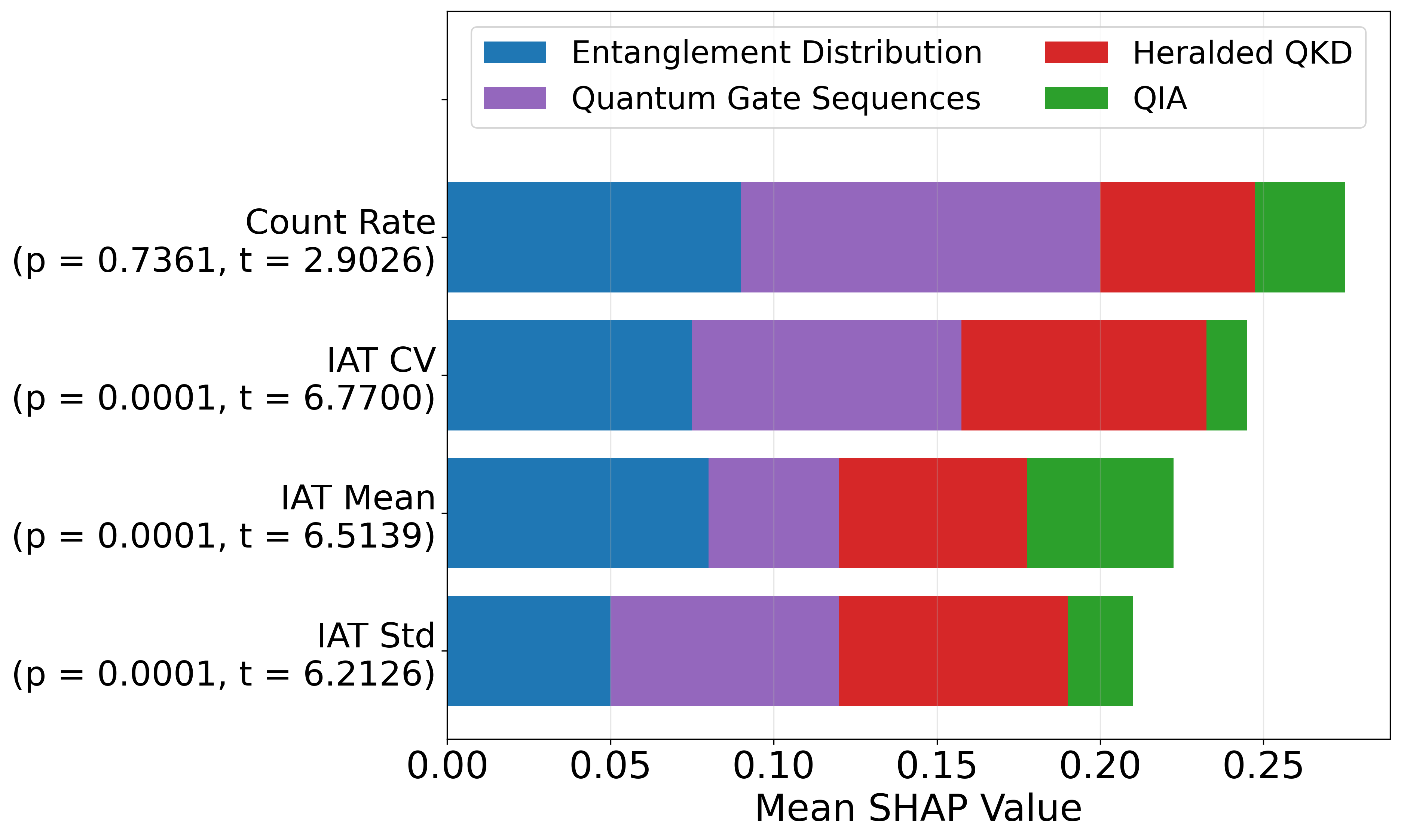}\label{shap_3070_tagger}}
\subfloat[]{\includegraphics[width=3.6in]{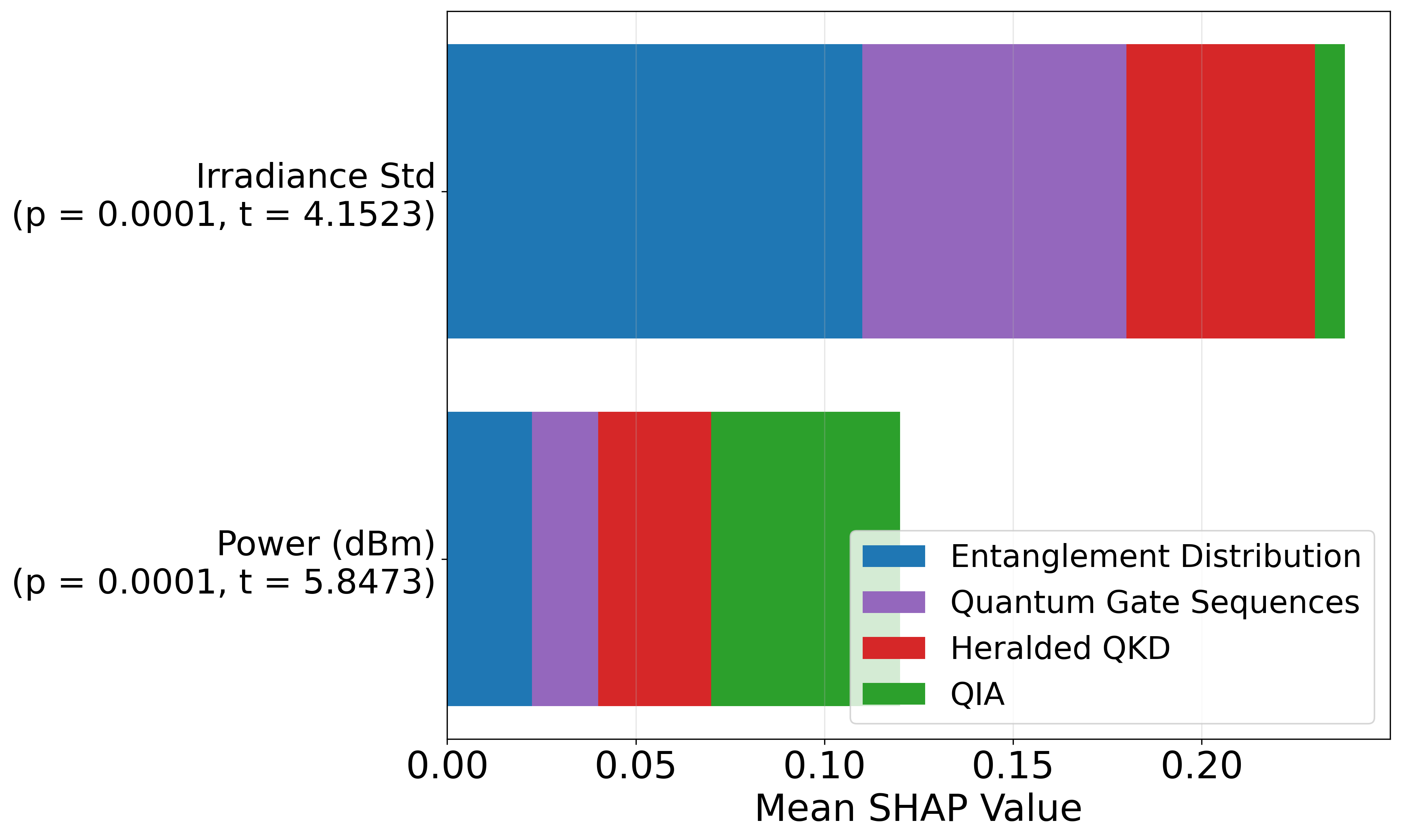}\label{shap_3070_power}}
\\
\subfloat[]{\includegraphics[width=3.6in]{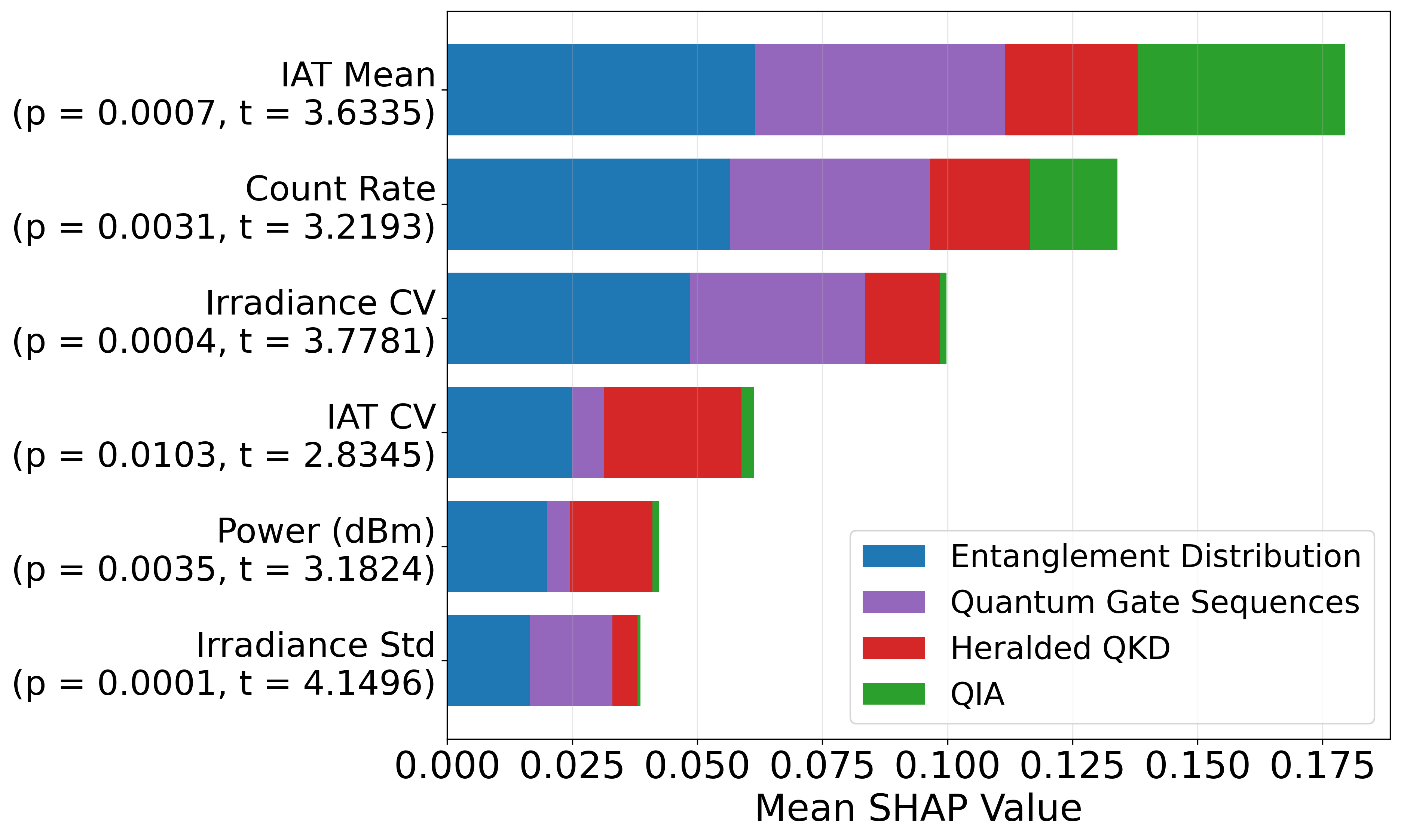}\label{shap_1090_combined}}
\subfloat[]{\includegraphics[width=3.6in]{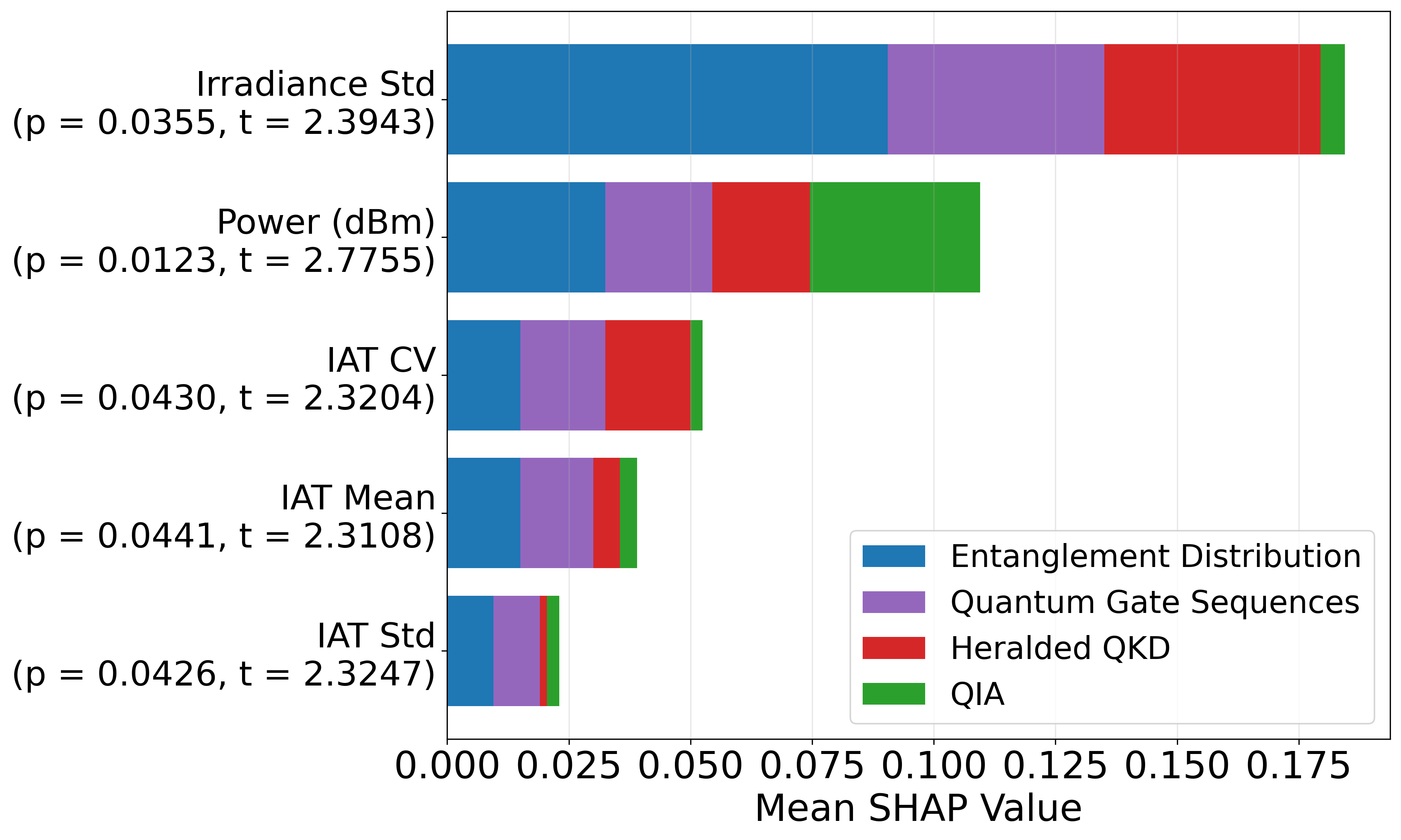}\label{shap_3070_combined}}
\caption{SHAP feature importance analysis for different sampling configurations and feature sets: (a) $10{:}90$ time tagger, (b) $10{:}90$ power meter, (c) $30{:}70$ time tagger, (d) $30{:}70$ power meter, (e) $10{:}90$ combined, and (f) $30{:}70$ combined.}
\label{fig:shap}
\end{figure*}

To interpret the behavior of the trained models, we analyze feature importance using SHAP values for the best performing model, the Bi-Stack LSTM, across different sampling configurations and feature sets, as shown in Fig.~\ref{fig:shap}. For clarity and interpretability, only the most influential features are shown in each SHAP plot. Feature selection is guided by the relative importance of SHAP values, where features with negligible contribution are omitted, resulting in a varying number of displayed features across configurations.
Figs.~\ref{shap_1090_tagger} and \ref{shap_3070_tagger} show the SHAP summaries for time tagger features under the $10{:}90$ and $30{:}70$ configurations, respectively. In both cases, timing features, particularly photon count rate and interarrival time statistics, consistently exhibit the highest contribution to classification decisions, indicating that timing dynamics are the primary source of protocol-specific information. Figs.~\ref{shap_1090_power} and \ref{shap_3070_power} illustrate the corresponding SHAP summaries for power meter features. Under the $10{:}90$ configuration, power-based features show limited contribution and weaker discriminative capability. In contrast, under the $30{:}70$ configuration, their contribution increases, with features such as irradiance variation and optical power becoming more informative. Figs.~\ref{shap_1090_combined} and \ref{shap_3070_combined} present the SHAP summaries for the combined feature set. These results show that timing features remain dominant, while power-based features provide complementary contributions that enhance classification performance. This explains the improved accuracy observed when combining both feature sets, particularly at higher sampling ratios. It is important to note that the set of dominant features varies across sampling configurations. This is expected, as SHAP values reflect the features that the model relies on for classification under a given data distribution. At lower sampling ratios, where the signal is weaker, feature importance is more distributed and less pronounced. In contrast, at higher sampling ratios, clearer patterns emerge, allowing the model to rely on a smaller subset of more informative features. Overall, these results confirm that timing photon statistics reveal protocol-level information, while intensity-based analysis provides complementary context that improves model performance.


\subsection{Discussion}

The results demonstrate that protocol fingerprinting in quantum communication systems is both feasible and robust under realistic observation constraints. Different protocols induce distinct temporal and statistical patterns at the physical layer, which can be captured through indirect measurements without accessing the underlying quantum states. Importantly, the coincidence analysis confirms that passive sampling does not break quantum correlations, as the measured S-values remain above the classical bound under all considered sampling configurations. This indicates that protocol fingerprinting can be performed without disrupting the execution of the underlying quantum communication protocols. At the same time, the analysis reveals a clear tradeoff between observability and coincidence rate. Higher sampling ratios provide richer information for classification but introduce greater reductions in photon count and coincidence rates. In contrast, lower sampling ratios significantly reduce coincidence rate while still preserving sufficient structure for reliable protocol distinguishability. The confusion matrix analysis further shows that classification errors are structured rather than random, reflecting similarities in the physical layer signatures of certain protocols. These misclassifications decrease as sampling increases and when complementary feature sets are used, indicating improved class separability. Overall, timing features derived from photon detection events provide stronger discriminative capability than intensity-based measurements, while combining both modalities yields the best performance. These findings reveal a previously underexplored information leakage channel in quantum networks, where protocol identity can be inferred through passive physical layer observations without breaking entanglement. The persistence of distinguishability under limited sampling highlights the practical relevance of this side channel in real-world quantum communication settings.

\section{Conclusion and Future Work}
\label{sec:conclusion}
In this paper, we experimentally investigated the feasibility of protocol fingerprinting in quantum communication networks through passive physical layer observations. Using a polarization-entangled photon testbed, we demonstrated that different quantum communication protocols induce distinct temporal and statistical patterns that can be captured without direct measurement of the underlying quantum states. Our results show that protocol identity can be inferred with high confidence under varying observation conditions. Higher sampling ratios improve classification performance by providing richer signal information, while lower sampling ratios still preserve sufficient structure for reliable protocol distinguishability. In addition, timing features derived from photon detection events provide stronger discriminative capability than intensity-based measurements, while the combination of both feature types yields the best overall performance. These findings reveal a previously underexplored information leakage channel in quantum networks, where protocol-level behavior can be inferred through passive observation. This challenges conventional assumptions about protocol-level privacy in quantum communication systems and highlights the need for new mechanisms to mitigate physical layer side-channel leakage.

Future work will extend this analysis to environments with increased noise at the observer, including attenuation, background light, and detector imperfections, as well as environmental variability and dynamic network conditions. Also, designing defense mechanisms that reduce or obfuscate protocol-specific physical layer signatures remains a critical direction for securing future quantum communication networks.

\bibliographystyle{IEEEtran}
\bibliography{References}
\end{document}